\documentclass[aps,prl,letterpaper,10pt,twocolumn,showpacs,superscriptaddress]{revtex4-1}
\usepackage{hyperref}
\usepackage{amssymb}
\usepackage{amsmath}
\usepackage{graphicx}
\usepackage{stmaryrd}
\usepackage{subfigure}
\usepackage{amsfonts}
\usepackage{color}
\newcommand{\wc}{\omega_{\rm c}}
\newcommand{\kc}{\kappa_{\rm c}}
\newcommand{\wm}{\Omega_{\rm m}}
\newcommand{\Gm}{\Gamma_{\rm m}}
\newcommand{\wt}{\omega_{\rm t}}
\newcommand{\gt}{\gamma_{\rm t}}
\newcommand{\xzpm}{x_{\rm{zpm}}}

\newcommand{\calD}{\mathcal{D}}

\newcommand{\Echr}{E_{\rm C}}
\newcommand{\Ejos}{E_{\rm J}}
\newcommand{\Dlasr}{\mathcal{E}_{\rm L}}

\newcommand{\mean}[1]{\overline{#1}}
\newcommand{\ket}[1]{|#1\rangle}

\newcommand{\ah}{\hat{a}}
\newcommand{\ahd}{\hat{a}^{\dagger}}
\newcommand{\bh}{\hat{b}}
\newcommand{\bhd}{\hat{b}^{\dagger}}
\newcommand{\ch}{\hat{c}}
\newcommand{\chd}{\hat{c}^{\dagger}}
\newcommand{\ph}{\hat{p}}
\newcommand{\phd}{\hat{p}^{\dagger}}

\newcommand{\Hh}{\hat{H}}

\begin{document}
\title{Quantum State Engineering with Circuit Electromechanical Three-Body Interactions}

\author{Mehdi Abdi}
\affiliation{Technische Universit{\"a}t M{\"u}nchen, Physik Department, James-Franck-Stra{\ss}e, 85748 Garching, Germany}
\author{Matthias Pernpeintner}
\affiliation{Technische Universit{\"a}t M{\"u}nchen, Physik Department, James-Franck-Stra{\ss}e, 85748 Garching, Germany}
\affiliation{Walther-Mei{\ss}ner-Institut, Bayerische Akademie der Wissenschaften, Walther-Mei{\ss}ner-Stra{\ss}e 8, 85748 Garching, Germany}
\affiliation{Nanosystems Initiative Munich, Schellingstra{\ss}e 4, 80799 M{\"u}nchen, Germany}
\author{Rudolf Gross}
\affiliation{Technische Universit{\"a}t M{\"u}nchen, Physik Department, James-Franck-Stra{\ss}e, 85748 Garching, Germany}
\affiliation{Walther-Mei{\ss}ner-Institut, Bayerische Akademie der Wissenschaften, Walther-Mei{\ss}ner-Stra{\ss}e 8, 85748 Garching, Germany}
\affiliation{Nanosystems Initiative Munich, Schellingstra{\ss}e 4, 80799 M{\"u}nchen, Germany}
\author{Hans Huebl}
\affiliation{Walther-Mei{\ss}ner-Institut, Bayerische Akademie der Wissenschaften, Walther-Mei{\ss}ner-Stra{\ss}e 8, 85748 Garching, Germany}
\affiliation{Nanosystems Initiative Munich, Schellingstra{\ss}e 4, 80799 M{\"u}nchen, Germany}
\author{Michael J. Hartmann}
\affiliation{Institute of Photonics and Quantum Sciences, Heriot-Watt University, Edinburgh, EH14 4AS, United Kingdom}

\date{\today}
\begin{abstract}
We propose a hybrid system with quantum mechanical three-body interactions between photons, phonons, and qubit excitations.
These interactions take place in a circuit quantum electrodynamical architecture with a superconducting microwave resonator coupled to a transmon qubit whose shunt capacitance is free to mechanically oscillate.
We show that this system design features a three-mode polariton--mechanical mode and a nonlinear transmon--mechanical mode interaction in the strong coupling regime.
Together with the strong resonator--transmon interaction, these properties provide intriguing opportunities for manipulations of this hybrid quantum system.
We show, in particular, the feasibility of cooling the mechanical motion down to its ground state and preparing various nonclassical states including mechanical Fock and cat states and hybrid tripartite entangled states.
\end{abstract}

%\pacs{}
\maketitle

%
%
%----------INTRODUCTION----------%
%\textit{Introduction}.
Quantum control of macroscopic objects is of great fundamental importance \cite{Caldeira1981} and
massive mechanical resonators strongly interacting with well-controlled quantum systems, e.g. photons and atomic excitations, are desired candidates for this purpose \cite{Monroe1996,Mancini1997,Knobel2003}.
These can be employed for preparing non-classical states in mechanical resonators \cite{Armour2002,Marshall2003,Abdi2012,Rips2012,Rips2013}, but are also of technological interest,
%Moreover, devices with a mechanical resonator strongly coupled to other quantum systems are also of technical interest,
e.g. for
weak force sensing \cite{Moser2013} and transduction of quantum information in quantum networks \cite{Rabl2010}.
Particularly, nonlinear quantum phenomena are very desirable for the above purposes as they considerably extend the options for manipulation and control of quantum systems.
Introducing an anharmonic part into a setup, can for example strengthen its couplings and enriches its physics via the nonlinearities \cite{Nigg2012,Ramos2013,Joeckel2014,Pirkkalainen2014,Pechal2014}.
Here, we propose a circuit electromechanical hybrid architecture that combines a nano-mechanical degree of freedom with both, an intrinsically nonlinear component in the form of a superconducting qubit and nonlinear interactions between the mechanical mode, the qubit excitations, and a harmonic mode of an electrical resonator.
As a key novelty, this architecture features three-body interactions between the mechanical and two electrodynamical degrees of freedom, which can not be approximated by effective two-body interactions due to the involved nonlinearities.

We explore a circuit quantum electrodynamical system consisting of a transmon qubit strongly coupled to a superconducting microwave resonator. In addition, the resonator--transmon system interacts with a nanomechanical oscillator.
The advantage of a transmon is its robustness against fluctuations of background charges achieved by increasing the ratio of Josephson and charging energies $\Ejos/\Echr$ \cite{Majer2007} at the cost of a reduced anharmonicity.
Nonetheless its nonlinearity can still be exploited for controllably producing single photons in a superconducting transmission line resonator via excitation exchange \cite{Hofheinz2008} or for controllably producing propagating surface acoustic phonons \cite{Gustafsson2014}.
Moreover, Josephson junctions integrated into a circuit electromechanical device can enhance its optomechanical couplings~\cite{Pirkkalainen2013,Heikkila2014,Rimberg2014,Pirkkalainen2014}.

\begin{figure}
\includegraphics[width=\linewidth]{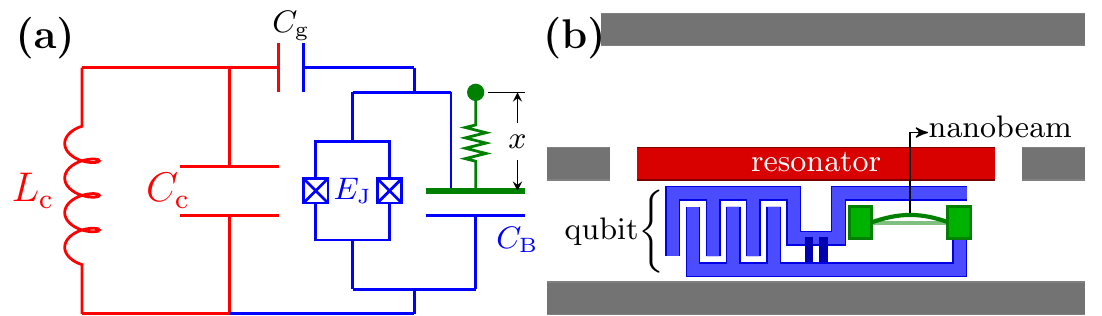}
\caption{(color online).
(a) Circuit diagram and (b) sketch of the hybrid system composed of a transmon qubit, a superconducting coplanar waveguide resonator and a mechanical oscillator.
}
\label{fig:scheme}
\end{figure}

Here, we propose to couple the transmon--cavity system to a mechanical resonator by replacing one of the transmon's shunt capacitor legs with an oscillating nano-beam, c.f. Fig.~\ref{fig:scheme}.
Hereby, we introduce a nonlinear coupling between the qubit and the mechanical resonator that can reach the strong coupling regime, i.e., the bare coupling rate can exceed the relaxation rates of the system.
In particular, as the nano-mechanical transmon is embedded in a microwave cavity the hybrid system features an electromechanical three-body interaction with a flux tunable coupling rate.
To show some assets of the system, we exploit these couplings and the anharmonicity of the transmon qubit to cool down the system to its ground state and prepare it in nonclassical states such as mechanical Fock and Schr{\"o}dinger cat states.

%
%
%----------MODEL----------%
\textit{The model}.
Our system is composed of a superconducting coplanar waveguide resonator equivalent to an $L_{\rm c}C_{\rm c}$ oscillator, capacitively coupled to a transmon qubit via a gate capacitance $C_{\rm g}$.
The shunt capacitance $C_{\rm B}$ of the transmon qubit depends on the position of a mechanical resonator as depicted in Fig.~\ref{fig:scheme}(b).
The Hamiltonian of the complete system can be written as ($\hbar=1$) \cite{supmat}
\begin{eqnarray}\label{genuine}
\Hh &=& \Hh_{0}+\Hh_{1}+\Hh_{\rm d}, \\
\Hh_{0} &=&\wt\ahd\ah -\lambda(\ahd)^{2}\ah^{2}
+\wm\bhd\bh +\wc\chd\ch +i\chi(\ah\chd -\ahd\ch), \nonumber\\
\Hh_{1} &=& \big[g_{\rm t}\ahd\ah +ig_{\rm t\rm c}(\ah\chd -\ahd\ch)\big](\bh+\bhd), \nonumber\\
\Hh_{\rm d} &=& \Dlasr(\ch e^{i\omega_{\rm L}t}+\chd e^{-i\omega_{\rm L}t}). \nonumber
\end{eqnarray}
Here, $\wt=\Echr(\sqrt{8\zeta}-1)$ with $\zeta = \Ejos/\Echr$ is the transition frequency between the ground and the first excited state of the transmon, which is modelled as an anharmonic oscillator with annihilation (creation) operator $\ah$ ($\ahd$) and Duffing nonlinearity $\lambda = \Echr/2$.
The charging energy of the qubit is $\Echr = e^{2}/2C_{\Sigma}$ with $C_{\Sigma}=C_{\rm g}+C_{\rm B}+C_{\rm J}$ ($C_{\rm J}$ is the capacitance of the Josephson junction).
The nano-mechanical resonator with natural frequency $\wm$ is described by phononic operators $\bh$ ($\bhd$) and its displacement is given by $\hat{x}=\xzpm(\bh+\bhd)$, where $\xzpm=\sqrt{\hbar/2m\wm}$ is its zero-point motion amplitude and $m$ its effective mass.
The microwave cavity oscillates with frequency $\wc$ and is characterized by the bosonic mode operators $\ch$ and $\chd$.
An external microwave field drives the cavity with amplitude $\Dlasr$.
The rate at which the qubit couples to the transmission line is $\chi=4\Echr n_{\rm{ac}}(\zeta/2)^{\frac{1}{4}}$ with $n_{\rm{ac}}$ the rms number of cavity induced Cooper pairs.
Moreover, the interactions between the mechanical resonator and the other parts of the system are quantified by the coupling rates $g_{\rm t}=g_{0}\sqrt{2\zeta}$ for the transmon--mechanical mode interaction and $g_{\rm t\rm c}=4g_{0}n_{\rm{ac}}(\zeta/2)^{\frac{1}{4}}$ for the \textit{three-body electro-mechanical mode interaction}, where $g_{0}$ is the bare coupling constant \cite{supmat}.
%\MakeUppercase{
Notably, both couplings are enhanced as the ratio $\Ejos/\Echr$ increases, which can be tuned in situ and in addition has the beneficial effect of enhancing the coherence time of the transmon.
%The two couplings already express the enhancement made on the bare coupling rate $g_{0}$ by the transmon qubit and that they can be tuned both via sample design and in situ by $\zeta$.
%Also, note that higher values of $\zeta$ desirably increases both stronger coupling rates and causes more coherent behaviour of the transmon.
%}
Finally, we mention that a rotating wave approximation (RWA) is applied to get the Hamiltonian $\Hh$. This is valid for $\zeta \gg 1$ and $\chi,g_{\rm t},g_{\rm t\rm c} \ll \wt,\wc$, which is compatible with the operation regime of our hybrid system.

In addition to the coherent evolution described by the above Hamiltonian, the system is affected by dissipation.
The energy relaxation rate of the transmon is $\gamma_{\rm t}=1/T_{1}$ and its total dephasing rate is $1/T_{2}^{*}=1/(2T_{1})+1/T_{\phi}$ where $\gamma_{\phi}= 1/T_{\phi}$ is the pure dephasing rate.
High quality superconducting qubits can be fabricated with relaxation and dephasing times as high as $T_{1}\approx 50~\mu$s and $T_{2}^{*}\approx 20~\mu$s \cite{Barends2013}.
Yet, even higher values, $T_{1}\approx 70 ~\mu$s and $T_{2}^{*}\approx 90 ~\mu$s, have been realized for 3D cavity setups \cite{Paik2011,Rigetti2012}.
In addition, the microwave photons of the cavity are subject to loss at a decay rate of $\kappa_{\rm c}$ and
the mechanical resonator is coupled to a thermal bath with the rate $\Gamma_{\rm m}=\wm/Q_{\rm m}$, where $Q_{\rm m}$ is its mechanical quality factor.
These dissipation processes are captured by a Liouvillian in Lindblad form.
Thus, the full dynamics of our system is described by the master equation
\begin{eqnarray}
\dot{\rho} &=& -i[\Hh,\rho] +\gt\calD_{\ah}\rho+\gamma_{\phi}\calD_{\ahd\ah}\rho +(\mean{n}+1)\Gm\calD_{\bh}\rho \nonumber\\
&&+\mean{n}\Gm\calD_{\bhd}\rho +\kc\calD_{\ch}\rho,
\label{master}
\end{eqnarray}
where $\calD_{\hat{o}}\rho=\hat{o}\rho\hat{o}^{\dagger}-(\hat{o}^{\dagger}\hat{o}\rho+\rho\hat{o}^{\dagger}\hat{o})/2$ is the dissipator and $\mean{n}=\big[\exp(\frac{\hbar\wm}{k_{\rm B}T})-1\big]^{-1}$ is the thermal phonon number.
%For the transmon, there is also a pure dephasing described by the second term.

%
%
%----------POLARITON----------%
\textit{Polariton-mechanical mode interaction}.
For typical configurations, the transmon--cavity interaction is in the strong coupling regime ($\chi \gg \gt,\kc$). It is therefore convenient to describe the subsystem of cavity photons and qubit excitations in terms of dressed state excitations called polaritons, which decouple the interaction $i\chi(\ah\chd -\ahd\ch)$.
In terms of these \textit{polaritonic} modes $\hat{p}_{\pm} =\alpha_{\pm} \ah \mp i\alpha_{\mp} \ch$ (the explicit forms of $\alpha_{\pm}$ are given in \cite{supmat}), the Hamiltonian (\ref{genuine}) reads
\begin{subequations}
\begin{eqnarray}
\Hh_{0} &=& \sum_{k=\pm}\Big[\omega_{k}\ph_{k}^{\dagger}\ph_{k} -\lambda_{k}(\ph_{k}^{\dagger})^{2}\ph_{k}^{2}\Big] +\Hh_{+-} +\wm\bhd\bh, \\
\Hh_{1} &=& \Big[\sum_{k=\pm}g_{k}\ph_{k}^{\dagger}\ph_{k}
+G(\ph_{+}^{\dagger}\ph_{-}+\ph_{+}\ph_{-}^{\dagger})\Big](\bh+\bhd),
\label{polariton}
\end{eqnarray}
\end{subequations}
where the polariton resonances are given by $\omega_{\pm}=\alpha_{\pm}^{2}\wt+\alpha_{\mp}^{2}\wc \pm 2\alpha_{+}\alpha_{-}\chi$, while the \textit{polariton--mechanical mode} coupling rates are $g_{\pm}=\alpha_{\pm}^{2}g_{\rm t}+2\alpha_{+}\alpha_{-} g_{\rm t\rm c}$ and $G=\alpha_{+}\alpha_{-} g_{\rm t}+(\alpha_{+}^{2}-\alpha_{-}^{2})g_{\rm t\rm c}$, and
the nonlinearity for each polariton is $\lambda_{\pm}=\alpha_{\pm}^{4}\lambda$ (Note that $G > g_{\rm t\rm c}$ due to the contribution from $g_{\rm t}$). There are thus three-body interactions for which the nonlinearity $\lambda$ precludes the linearization of the terms $\phd_{+}\ph_{-}$ or $\ph_{+}\phd_{-}$. This is in contrast to the original three-mode interaction with strength $g_{\rm t\rm c}$ or to optomechanical couplings in standard settings \cite{Loerch2015}.
%In addition to the original three-body interactions with strength $g_{\rm t\rm c}$, there are thus three-body interactions that emerge 
%due to the strong qubit-cavity coupling $\chi$ (responsible for the emergence of two polaritons modes) in combination with the qubit nonlinearity $\lambda$ (precluding linearizations of the terms $\ph_{+}^{\dagger}\ph_{-}$ or $\ph_{+}\ph_{-}^{\dagger}$).
We have included all the inter-polariton interaction terms in $\Hh_{+-}$, see \cite{supmat}. Note that except for an intensity--intensity interaction, all these inter-polariton interactions can be neglected in a RWA \cite{supmat} provided $\omega_{+}-\omega_{-}=[\Delta^{2}+4\chi^{2}]^{\frac{1}{2}} \gg \lambda$, where $\Delta=\wt-\wc$ is the detuning between the transmon and cavity frequencies.
This condition calls for a large photon--qubit coupling rate ($\chi \gg \lambda$) and/or large transmon--cavity detuning ($\Delta \gg \lambda$). To ensure the validity of the above RWA, we work in the off-resonance regime with $\Delta \gg \lambda$.
The polariton--mechanical interactions in Eq.~(\ref{polariton}) 
provide us with a toolbox for the quantum control of the state of the mechanical resonator, which is our main interest for the proposed architecture in this letter.

%
%
%----------COOLING----------%
\textit{Cooling the mechanical resonator}.
A first question of interest is whether our setup allows for ground state cooling of the mechanical mode as this is a prerequisite for many state preparation protocols.
We show that this is indeed feasible using sideband cooling \cite{Teufel2011}.
In principle, both interactions of (\ref{polariton}) are capable of performing the task.
For the three-mode interaction one would transfer a mechanical phonon and a lower polariton excitation into a higher polariton excitation, $\omega_{-}+\wm = \omega_{+}$, which subsequently decays.
However, a simpler and more efficient route is to use the couplings $\phd_{\pm}\ph_{\pm}(\bh +\bhd)$ at large cavity--transmon detunings.
Here, one polariton is dominantly photon-like, while the other describes mainly a transmon excitation.
In this regime, the photon-like polariton is practically decoupled from the mechanical mode, while the transmon-like polariton strongly interacts with it at a coupling rate close to $g_{\rm t}$.
For this kind of interaction, the final occupation number of the mechanical mode is limited by the total dephasing time $T_{2}^{*}$ of the qubit and the ground state can be reached for $\wm T_{2}^{*}>1$ \cite{Rabl2010a}.

\begin{figure}
\includegraphics[width=\linewidth]{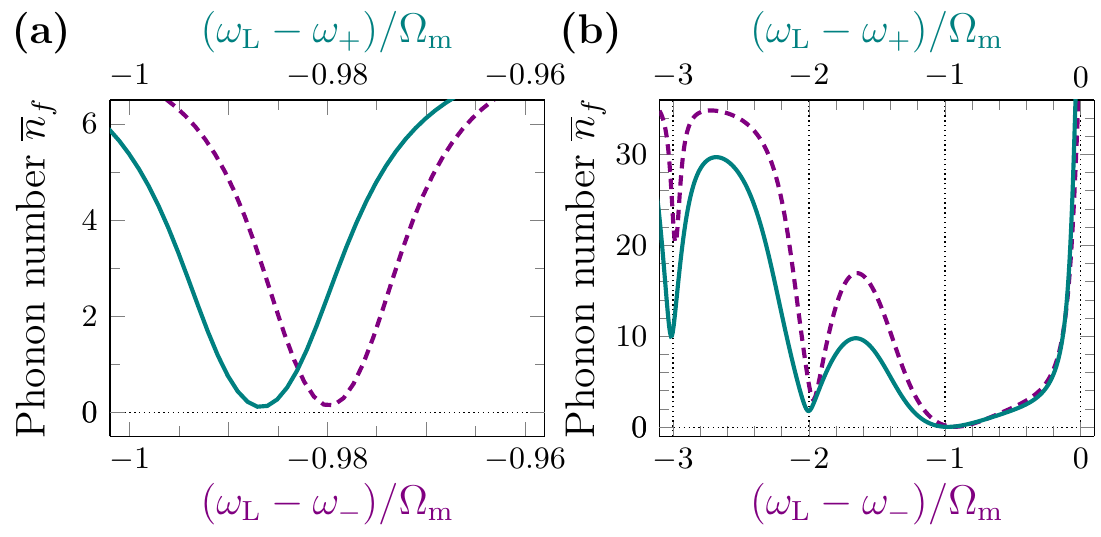}
\caption{(color online).
Mechanical final phonon numbers versus drive detuning when only one of the polaritons $\ph_{+}$ (solid lines) or $\ph_{-}$ (dashed lines) is driven:
(a) parameter set\#1 with $\zeta=150 \pm 50$ and (b) parameter set\#2 with $\zeta=142 \pm 60$.
}
\label{fig:cooling}
\end{figure}

We numerically solve Eq.~(\ref{master}) with Hamiltonian (\ref{genuine}) for two different sets of parameters.
Set\#1: $\zeta_{\rm c} = 150$ (the resonance value, i.e., $\wt(\zeta_{\rm c})=\wc$), $m = 1$ pg, $g_{0}=18.2$ kHz, $\wm/2\pi = 10$ MHz, $\kc/2\pi = 10$ kHz, $\gt/2\pi = 3$ kHz, and $n_{\rm{ac}} = 8.5\times 10^{-4}$.
Set\#2: $\zeta_{\rm c} = 142$, $m = 3$ pg, $g_{0}=20.6$ kHz, $\wm/2\pi = 1$ MHz, $\kc/2\pi = 50$ kHz, $\gt/2\pi = 5$ kHz, and $n_{\rm{ac}} = 1.4\times 10^{-2}$.
We also take $\gamma_{\phi} = 2\gt$ and use the common parameters $\Echr/2\pi = 0.5$~GHz, $\wc/2\pi \approx 17$~GHz, and $Q_{\rm m} = 10^{6}$.
The considered mechanical parameters are compatible with experimental reported values \cite{Regal2008,Zhou2013,Pernpeintner2014}.
In Fig.~\ref{fig:cooling} we plot the numerical results for the final phonon numbers achievable by cooling via either of the polaritons.
We find that it is possible to cool the mechanical resonator from $\mean{n} \approx 20$ phonons (corresponding to an ambient temperature $T \simeq 10$ mK) to $\mean{n}_{f} \approx 0.12$ for the parameter set\#1 and from $\mean{n} \approx 100$ phonons (environment temperature $T\simeq 5$ mK) to $\mean{n}_{f} \approx 0.03$ for the parameter set\#2.
The multiple cooling resonances apparent in Fig.~\ref{fig:cooling}(b) are a signature of the nonlinearity of the coupling \cite{Nunnenkamp2012} and thus provide a measurable witness for the nonlinearity of the transmon--viberational mode interaction.
Having shown the feasibility of ground state cooling, we now describe two state preparation protocols enabled by our device.

%
%
%----------FOCK----------%
\textit{Mechanical Fock states}.
We first describe a protocol for preparing the mechanical resonator in Fock states.
Our strategy here is to first generate individual polaritonic excitations and then transfer them to the mechanics via the three-mode interaction in Eq. (\ref{polariton}).
To this end, sideband cooling first brings the mechanical resonator close to its ground state.
%(causing a small increment in the transmon excitation probability).
Properly shaped microwave pulses at suitable frequencies can generate single-qubit rotations for the polaritons \cite{Steffen2003}.
Hence, the polariton with higher frequency is excited by such a pulse.
Then the transmon is tuned to the point where $\omega_{+}-\omega_{-}=\wm$ and the system evolves for a time $\tau_{1}=\pi/2G$, which converts the higher energy polariton into a lower energy polariton and a single phonon in the mechanical resonator.
The generated lower energy polariton is finally annihilated by another microwave pulse, leaving the system in a single phonon Fock state.

Strong three-body interactions and hence fast excitation transfer could of course be achieved for $\wt \approx \wc$ \cite{supmat}.
Yet, in this regime all polariton--mechanical interactions have the same strength $g_{+}=g_{-}=G \approx g_{\rm t}/2$, which enables additional undesirable transfer channels that hamper the protocol.
Hence, we demand for sufficiently large mechanical frequencies to suppress these unwanted interactions via a RWA and simultaneously ensure $\omega_{+}-\omega_{-}=\wm$.
At the same level of accuracy the three-mode interaction becomes $G(\ph_{+}^{\dagger}\ph_{-}\bh+\ph_{+}\ph_{-}^{\dagger}\bhd)$.
Furthermore, the transfer process must be much faster than the decoherence rates of the system, therefore, the restrictions on the system are $\max \{\gamma_{\rm t},\tilde{\Gamma}_{\rm m}\} \ll g_{\rm t} \ll \wm$, where $\tilde{\Gamma}_{\rm m}\approx\mean{n}\Gamma_{\rm m}$ is the effective mechanical decoherence rate.
The parameter set\#1 satisfies the above criteria.

\begin{figure}
\includegraphics[width=\linewidth]{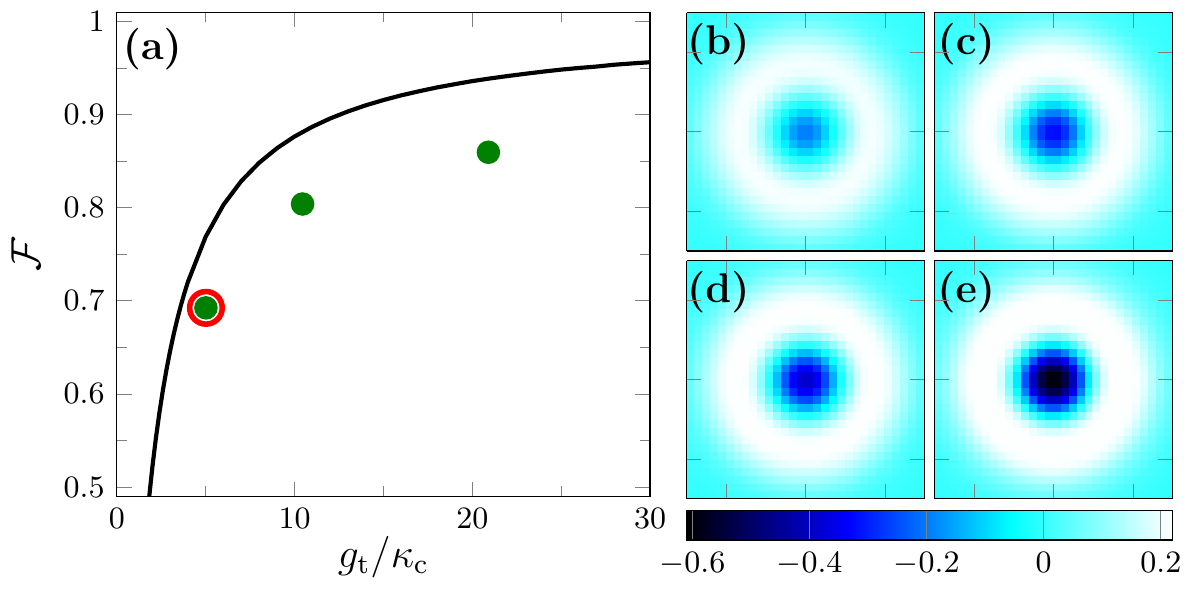}
\caption{(color online).
(a) Fidelity of prepared single phonon state as a function of the transmon--mechanical coupling rate; for ideal initial ground state (solid line) and attainable ground state (dots).
The point marked by the red circle corresponds to the parameter set\#1.
(b)--(e) Wigner quasi-probability distribution of the prepared Fock states with $g_{\rm t}/\kc \approx$ (b) 5, (c) 10, and (d) 21.
(e) Ideal single phonon state, for comparison. 
}
\label{fig:fock}
\end{figure}

To provide evidence for its feasibility, we numerically simulate the protocol, including the initial cooling to the ground state, by solving the full master equation (\ref{master}).
As very high fidelities for single-qubit gates have already been demonstrated, we neglect errors in the polariton excitation and de-excitation steps.
The fidelity for a single-phonon state prepared by sideband cooling followed by the above protocol reaches 70\% for the parameter set\#1, see Fig.~\ref{fig:fock}, and
could even be enhanced further by increasing the coupling rate and/or starting from better ground states, e.g. by employing qubit reset methods \cite{Geerlings2013}.
%and stiffening the mechanical resonator by applying inhomogeneous electric fields during the cooling process \cite{Rips2012,Rips2013}.
To achieve higher number states, the process can be repeated until the target state is reached, where the interaction times  for the swap need to be adjusted to $\tau_{n}=\pi/2\sqrt{n}G$ with $n$ being the number of mechanical phonons which will be obtained at the end of each stage.
Note that, although the inter-polariton interactions are considerable, they will not play a role in this process as long as the total number of polaritons does not exceed unity \cite{supmat}.
Once prepared, the state of the mechanical resonator can be analyzed by adapting the measurement scheme pioneered in \cite{Hofheinz2008}, i.e. tuning the transmon to the point where $\omega_{+}-\omega_{-}=\wm$ for various interaction times and reading out its excitation probability.

We note that the existence of the cavity mode is crucial for the above protocol since the qubit--mechanical interaction is not in the form of a state transfer Hamiltonian, nor can it be changed to such form by intensely driving the (anharmonic) qubit
and the original three-mode interaction is not strong enough to be used in this way. 
Yet, the strong coupling of the cavity to the qubit generates a strong state transfer interaction of three-body form between the two polaritons on the one hand and the mechanical resonator on the other hand.

%
%
%----------ENTANGLED----------%
\textit{Tripartite hybrid entanglement}.
We now turn to propose a protocol for preparing a non-Gaussian tripartite entanglement between qubit, cavity, and the mechanical resonator in our system.
Such states are of interest both from fundamental and technical points of view \cite{Greenberger1989}.
In order to describe the steps for creating them, we go back to the original (not dressed) picture of the system and consider an effective three-level model for the transmon with ground state $\ket{0}_{\rm t}$ and excited states $\ket{1}_{\rm t}$ and $\ket{2}_{\rm t}$.

The mechanical resonator is first cooled down to its ground state with the transmon and cavity off-resonance.
By applying a $[\pi/2]_{0\leftrightarrow 1}$ pulse, the qubit is prepared in a symmetric superposition of ground and first excited state.
Then we let the mechanical resonator interact with the qubit.
As the force exerted on the nano-beam depends on the state of the transmon, such an interaction results in a conditional displacement of the mechanical system from the origin of the phase space.
The maximal amount of displacement is $2g_{\rm t}/\wm$, which is achieved when the interaction duration equals half the mechanical oscillation period.
However, to have two distinguishable peaks in the mechanical phase space one needs $g_{\rm t} \gtrsim \wm$, which is not the case in our system.
This hurdle can be circumvented by applying a sequence of $N_{\rm p}$ regularly spaced $[\pi]_{0\leftrightarrow 1}$ pulses to the qubit with time intervals equal to half of the mechanical period.
By choosing an odd number of pulses $N_{\rm p}$, apart from an irrelevant global phase factor, one arrives at $\frac{1}{\sqrt{2}}\big(\ket{0}_{\rm t}\ket{0}_{\rm c}\ket{\beta}_{\rm m}+\ket{1}_{\rm t}\ket{0}_{\rm c}\ket{-\beta}_{\rm m}\big)$ with $\beta=(N_{\rm p}+1)g_{\rm t}/\wm$ \cite{Tian2005,Asadian2014}.
In the next step, the mechanical resonator can be turned into a superposition of odd and even cat states, by applying a $[\pi/2]_{0\leftrightarrow 1}$ pulse to the transmon: $\ket{\Psi_{\rm{tm}}}=\frac{1}{2}\big(\ket{0}_{\rm t}\ket{0}_{\rm c}\ket{\psi_{+}}_{\rm m}+\ket{1}_{\rm t}\ket{0}_{\rm c}\ket{\psi_{-}}_{\rm m}\big)$, which is already a bipartite qubit--mechanical entangled state.
Here, $\mathcal{N}_{\pm}\ket{\psi_{\pm}}=\mathcal{N}_{\pm}\big(\ket{\beta}\pm \ket{-\beta}\big)$ with the normalization factor $\mathcal{N}_{\pm}=\big[2\pm 2 e^{-2|\beta|^{2}}\big]^{-\frac{1}{2}}$ is an even/odd cat state.
Now, a single photon in the cavity can be conditionally produced via a $[\pi]_{1\leftrightarrow 2}$ pulse that flips the qubit from its first to its second excited state $\ket{1}_{\rm t} \rightarrow \ket{2}_{\rm t}$.
Then a flux pulse of duration $\pi/2\sqrt{2}\chi$ sets this transmon transition in resonance with the cavity.
Therefore, the second qubit excitation is transferred into the cavity, leading to
\begin{equation}
\ket{\Psi_{\rm{tmc}}} = \frac{1}{2}\big(\ket{0}_{\rm t}\ket{0}_{\rm c}\ket{\psi_{+}}_{\rm m}+\ket{1}_{\rm t}\ket{1}_{\rm c}\ket{\psi_{-}}_{\rm m}\big).
\label{ghz}
\end{equation}
The state (\ref{ghz}) is a hybrid Greenberger--Horne--Zeilinger state \cite{Gerry1996}.
Evidently, $\ket{\Psi_{\rm{tmc}}}$ could also be reduced either to even or odd cat states of the mechanical mode by performing a post-selection based on the read out of the qubit or cavity state.

To attain macroscopically distinguishable mechanical cat states a minimum number of pulses $N_{\rm p}$ is required. 
On the other hand, the realizable $N_{\rm p}$ is limited by the decoherence rates of the system.
Thus, the parameter regime allowing for a successful preparation of the state (\ref{ghz}) is $\pi\times\max\{\gamma_{\rm t},\tilde{\Gamma}_{\rm m}\} < \frac{\wm}{N_{\rm p}} < g_{\rm t}$.
Note also that, employing the third level for cloning the qubit excitations as cavity photons is necessary to get the state (\ref{ghz}).
Furthermore, it is desirable to transfer the qubit second-excitations into the cavity fast enough to decrease the unwanted displacements, so we also demand $\chi \gg \wm$.
Since we need to work with the transmon--mechanical mode interaction, we chose off-resonance as the appropriate working regime, where the three-mode interaction is negligible.

\begin{figure}[t]
\includegraphics[width=\linewidth]{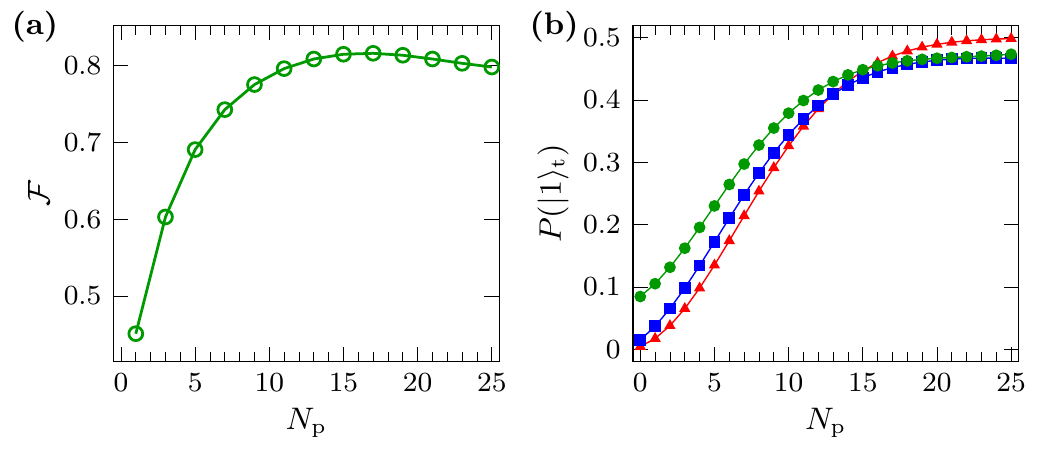}
\caption{(color online).
(a) Fidelity of the prepared odd cat state versus $N_{\rm p}$.
(b) Probability of finding the transmon in its first excited state for the target state $\ket{\Psi_{\rm{tmc}}}$ (red triangle), for the state resulting from a simulation starting from the ideal ground state (blue square), and a simulation starting from an attainable ground state (green circle).
The parameter set\#2 is used in these simulations.
}
\label{fig:ghz}
\end{figure}

In Fig.~\ref{fig:ghz}(a) the fidelity of the prepared odd cat state is plotted versus the number of applied pulses.
Signatures of its generation can be obtained %by performing a state tomography 
via extracting phonon number probabilities with the method exploited in Ref.~\cite{Hofheinz2009}.
The protocol itself can be verified in an easier way by measuring the probability of finding the qubit in its first excited state or detecting a single photon in the output of the cavity as a function of the number of pulses $N_{\rm p}$ applied to the qubit.
Theoretically, for the ideal state $\ket{\Psi_{\rm{tmc}}}$ these probabilities are $P(\ket{1}_{\rm t})=P(\ket{1}_{\rm c})=\frac{1}{2}\big(1-\exp{\{-2(N_{\rm p}+1)^{2}g_{\rm t}^{2}/\wm^{2}\}}\big)$.
Fig.~\ref{fig:ghz}(b) shows $P(\ket{1}_{\rm t})$ and the probability of finding the transmon in its first excited state (or detecting a single photon) for the states resulting from numerical simulations of the preparation protocol neglecting errors in single-qubit gate operations.
The coincidence of the simulations with the theoretical ideal curve confirms the feasibility of the protocol.

%
%
%----------CONCLUSION----------%
%\textit{Conclusion}.
%In conclusion, we have introduced a hybrid system, offering versatile quantum control over a mechanical resonator.
%This includes cooling the mechanical resonator to its ground state, preparing it in phononic number states, and macroscopically distinguishable cat states via post-selection of non-Gaussian hybrid entangled states.

\textit{Acknowledgements}
MA acknowledges support by the Alexander von Humboldt Foundation via a postdoctoral fellowship, HH and RG by the German Research Foundation (DFG) via the SFB 631 and MJH by the DFG via HA 5593/3-1 and the SFB 631.

%
%
%----------REFERENCES----------%
\bibliography{transmon}

\clearpage
\onecolumngrid
%\begin{widetext}
\begin{center}
\textbf{\large Supplemental Materials: Quantum State Engineering with Circuit Electromechanical Three-Body Interactions}
\end{center}

\setcounter{equation}{0}
\setcounter{figure}{0}
\setcounter{table}{0}
\setcounter{page}{1}
\makeatletter
\renewcommand{\theequation}{S\arabic{equation}}
\renewcommand{\thefigure}{S\arabic{figure}}
\renewcommand{\bibnumfmt}[1]{[S#1]}
\renewcommand{\citenumfont}[1]{S#1}

\section{The model}
The system we consider is composed of a coplanar waveguide microwave cavity capacitively coupled to a transmon qubit whose shunt capacitance depends on the position of a mechanical resonator.
The Hamiltonian of the system is given by
\begin{equation}
\Hh=4\Echr(\hat{x})(\hat{n}-n_{\rm g})^{2} -\Ejos\cos\hat{\varphi} +\hbar\wm\bhd\bh +\hbar\wc^{\prime}\chd\ch,
\label{full}
\end{equation}
where $\hat{n}$ and $\hat{\varphi}$ are the superconducting charge number and phase operators, satisfying the commutation relation $[\hat{\varphi},\hat{n}]=i$ and $n_{\rm g}$ is the offset charge of the device which contains both dc and ac contributions.
The mechanical resonator is characterized by its natural frequency $\wm$ and phonon annihilation (creation) operator $\bh$ ($\bhd$).
Its position is given by $\hat{x}=\xzpm(\bh+\bhd)$ where $\xzpm=\sqrt{\hbar/2m\wm}$ is the mechanical zero-point motion.
And $\wc^{\prime}$ is the `bare' cavity mode frequency represented by bosonic operator $\ch$.

For large ratios $\Ejos/\Echr \equiv \zeta$, the anharmonicity of a transmon is not very high and we describe it by expanding the cosine term around $\varphi =0$ and keep up to the fourth power in $\hat{\varphi}$,
\begin{equation}
\Hh =4\Echr(\hat{x})(\hat{n}-n_{\rm g})^{2} +\frac{\Ejos}{2}\hat{\varphi}^{2} -\frac{\Ejos}{24}\hat{\varphi}^{4}
+\hbar\wm\bhd\bh +\hbar\wc^{\prime}\chd\ch.
\end{equation}
The offset charge $n_{\rm g}$ is induced by an applied dc voltage, intrinsic defects and by the stripline resonators field.
The latter yields an ac component in $n_{\rm g}$, so we write $n_{\rm g}=n_{\rm{dc}}+n_{\rm{ac}}(\ch+\chd)$, where $n_{\rm{ac}}\equiv C_{\rm g}V_{0}/2e$ is the rms number of the vacuum induced Cooper pairs.
Here, $V_{0}=\sqrt{\hbar\wc^{\prime}/2C_{\rm c}}$ is the root mean square voltage of the resonator's vacuum field, $C_{\rm c}$ is its capacitance, and $C_{\rm g}$ the gate capacitance.
For a transmon qubit, the $n_{\rm dc}$ can be set equal to zero because its influence on the energy levels is negligible.
Then the Hamiltonian reads
\begin{eqnarray}
\Hh &=&4\Echr(\hat{x})\hat{n}^{2}+\frac{\Ejos}{2}\hat{\varphi}^{2}-\frac{\Ejos}{24}\hat{\varphi}^{4} \nonumber\\
&&+\hbar\wm\bhd\bh +\hbar\wc^{\prime}\chd\ch \nonumber\\
&&+4\Echr(\hat{x}) n_{\rm{ac}}^{2}(\ch+\chd)^{2} \nonumber\\
&&-8\Echr(\hat{x}) n_{\rm{ac}}\hat{n}(\ch+\chd).
\end{eqnarray}
We now Taylor expand the charging energy $\Echr$ and keep up to the first power of $\hat{x}$:
$$\Echr(\hat{x}) \approx \Echr +g_{0}(\bh+\bhd),$$
where we have defined a bare coupling constant $g_{0}=\frac{\Echr}{d_{0}}\frac{C_{\rm B}}{C_{\Sigma}}\xzpm$ with $d_{0}$ the equilibrium distance between the plates of the shunt capacitor with capacitance $C_{\mathrm{B}}$ and $C_{\Sigma}=C_{\rm g}+C_{\rm B}+C_{\rm J}$ the total capacitance.
Bosonic annihilation and creation operators can be defined for the transmon such that
\begin{eqnarray*}
\hat{n} &=& -\frac{i}{2}\bigg[\frac{\Ejos}{2\Echr}\bigg]^{\frac{1}{4}} (\ah -\ahd), \\
\hat{\varphi} &=& \bigg[\frac{2\Echr}{\Ejos}\bigg]^{\frac{1}{4}} (\ah +\ahd).
\end{eqnarray*}
The Hamiltonian can now be rewritten in terms of bosonic annihilation and creation operators
\begin{eqnarray}
\Hh &=&\hbar\wt^{\prime}\ahd\ah -\frac{\Echr}{12}(\ah+\ahd)^{4} \nonumber\\
&&+\hbar\wm\bhd\bh 
+\hbar\wc^{\prime}\chd\ch +4\Echr n_{\rm{ac}}^{2}(\ch+\chd)^{2} \nonumber\\
&&+i\hbar\chi(\ah-\ahd)(\ch+\chd) \nonumber\\
&&-\hbar \frac{g_{\rm t}}{2}(\ah-\ahd)^{2}(\bh+\bhd) \nonumber\\
&&+\hbar \frac{g_{\rm c}}{2}(\bh+\bhd)(\ch+\chd)^{2} \nonumber\\
&&+i\hbar g_{\rm t\rm c}(\ah-\ahd)(\bh+\bhd)(\ch+\ch ^{\dagger}) \nonumber\\
&&-i\hbar\Dlasr(\ch e^{i\omega_{\rm L}t}-\chd e^{-i\omega_{\rm L}t}),
\label{Sfull}
\end{eqnarray}
where $\hbar\wt^{\prime}=\sqrt{8\Ejos\Echr}$ is the frequency of oscillations between ground state and the first excited state of the transmon
and $\hbar\chi=4\Echr n_{\rm{ac}}(\zeta/2)^{\frac{1}{4}}$ is the qubit to transmission line coupling rate.
Also, the transmon-mechanical $\hbar g_{\rm t}=g_{0}\sqrt{2\zeta}$ and electro-mechanical $\hbar g_{\rm c}=8g_{0}n_{\rm{ac}}^{2}$ coupling rates and the three mode interaction rate as $\hbar g_{\rm t\rm c}=4g_{0}n_{\rm{ac}}(\zeta/2)^{\frac{1}{4}}$ have been defined.
The last line of the Hamiltonian includes the coherent drive of the system by a microwave input.
Since both $g_{0}$ and $n_{\rm{ac}}$ usually acquire very small values, the electromechanical interaction rate $g_{\rm c}$ is negligible in this setup.

After applying the rotating wave approximation which is valid when $\Echr \ll 12\hbar\wt^{\prime}$ (or equivalently when $\zeta \gg 1$) for the Duffing nonlinearity term and $\chi,g_{\rm t},g_{\rm c},g_{\rm t\rm c} \ll \wt^{\prime},\wc^{\prime}$ for the interactions; the Hamiltonian of the system reduces to Eq.~(1) in the main text with the effective cavity frequency $\wc=\wc^{\prime}+8\frac{\Echr}{\hbar}n_{\rm{ac}}^{2}$.
The parameters listed in Table~\ref{parameters} show that this approximation is valid indeed.

%
%
%----------POLARITON----------%
\section{Polariton-mechanical interaction}
Since the transmon--cavity interaction is in the strong coupling regime ($\chi \gg \gamma_{\rm t},\gamma_{\rm c}$) the subsystem of cavity photons and qubit excitations can be described in terms of polaritons.
This polaritonic picture can be useful for eliminating the major interaction term and diagonalizes some terms of the Hamiltonian.
The nonlinear part of the Hamiltonian can bring in polariton-polariton interactions which can spoil this simplified picture.
However, if the nonlinearity is much weaker than the photon-qubit coupling $\Echr \ll 2\hbar\chi$ and/or the transmon-cavity frequency difference $\Echr \ll 2\hbar\Delta$, we can neglect such inter-polariton interactions in a rotating wave approximation.
For now, we will keep all the terms for completeness, however, in some situations we will focus on this regime.
We define the following polaritonic modes
\begin{align*}
\ph_{+} &=\alpha_{+} \ah -i\alpha_{-} \ch, \\
\ph_{-} &=\alpha_{-} \ah +i\alpha_{+} \ch,
\end{align*}
where $\alpha_{+}$ and $\alpha_{-}$ are real positive numbers satisfying $\alpha_{+}^{2}+\alpha_{-}^{2}=1$ and given by
\begin{equation}
\alpha_{\pm} = \frac{1}{\sqrt{2}}\Big(1 \pm \frac{\Delta}{\sqrt{\Delta^{2}+4\chi^{2}}}\Big)^{\frac{1}{2}}.
\end{equation}
Finally, we arrive at the following polariton--mechanical Hamiltonian:
\begin{eqnarray}
\frac{\Hh}{\hbar} &=&\wm\bhd\bh
+\omega_{+}\ph_{+}^{\dagger}\ph_{+} -\lambda_{+}\ph_{+}^{\dagger}\ph_{+}^{\dagger}\ph_{+}\ph_{+} 
+\omega_{-}\ph_{-}^{\dagger}\ph_{-} -\lambda_{-}\ph_{-}^{\dagger}\ph_{-}^{\dagger}\ph_{-}\ph_{-}
+\frac{\Hh_{+-}}{\hbar} \nonumber\\
&&+(g_{+}\ph_{+}^{\dagger}\ph_{+}+g_{-}\ph_{-}^{\dagger}\ph_{-})(\bh+\bhd)
+G (\ph_{+}^{\dagger}\ph_{-}+\ph_{+}\ph_{-}^{\dagger})(\bh+\bhd) \nonumber\\
&&-\alpha_{-}\Dlasr(\ph_{+}e^{i\omega_{\rm L}t}+\ph_{+}^{\dagger}e^{-i\omega_{\rm L}t})
+\alpha_{+}\Dlasr(\ph_{-}e^{i\omega_{\rm L}t}+\ph_{-}^{\dagger}e^{-i\omega_{\rm L}t}).
\label{Spolariton}
\end{eqnarray}
Here, the following coupling constants have been introduced:
Conventional \textit{polaritomechanical} coupling rates $g_{+}=\alpha_{+}^{2}g_{\rm t}+\alpha_{-}^{2}g_{\rm c}+2\alpha_{+}\alpha_{-} g_{\rm t\rm c}$ and $g_{-}=\alpha_{-}^{2}g_{\rm t}+\alpha_{+}^{2}g_{\rm c}-2\alpha_{+}\alpha_{-} g_{\rm t\rm c}$, and the three-mode interaction $G =\alpha_{+}\alpha_{-} (g_{\rm t}-g_{\rm c})+(\alpha_{+}^{2}-\alpha_{-}^{2})g_{\rm t\rm c}$.
The nonlinearities of the polaritons are $\lambda_{+}=\Echr\alpha_{+}^{4}/2\hbar$ and $\lambda_{-}=\Echr\alpha_{-}^{4}/2\hbar$.
The inter-polariton interaction terms are included in $\Hh_{+-}$
\begin{eqnarray}
\Hh_{+-} &= -\dfrac{\Echr}{2}\Big[&
\alpha_{+}^{4}\ph_{+}^{\dagger}\ph_{+}^{\dagger}\ph_{+}\ph_{+} 
+\alpha_{-}^{4}\ph_{-}^{\dagger}\ph_{-}^{\dagger}\ph_{-}\ph_{-}
+4\alpha_{+}^{2}\alpha_{-}^{2}\ph_{+}^{\dagger}\ph_{+}\ph_{-}^{\dagger}\ph_{-} \nonumber\\
&&+\alpha_{+}^{2}\alpha_{-}^{2}(
+\ph_{+}^{\dagger}\ph_{+}^{\dagger}\ph_{-}\ph_{-}
+\ph_{+}\ph_{+}\ph_{-}^{\dagger}\ph_{-}^{\dagger}) \nonumber\\
&&+2\alpha_{+}^{3}\alpha_{-}(
\ph_{+}^{\dagger}\ph_{+}^{\dagger}\ph_{+}\ph_{-}
+\ph_{+}^{\dagger}\ph_{+}\ph_{+}\ph_{-}^{\dagger}) \nonumber\\
&&+2\alpha_{+}\alpha_{-}^{3}(
\ph_{+}\ph_{-}^{\dagger}\ph_{-}^{\dagger}\ph_{-}
+\ph_{+}^{\dagger}\ph_{-}^{\dagger}\ph_{-}\ph_{-})\Big].
\label{Spp}
\end{eqnarray}

\begin{figure}[t]
\includegraphics[width=0.4\linewidth]{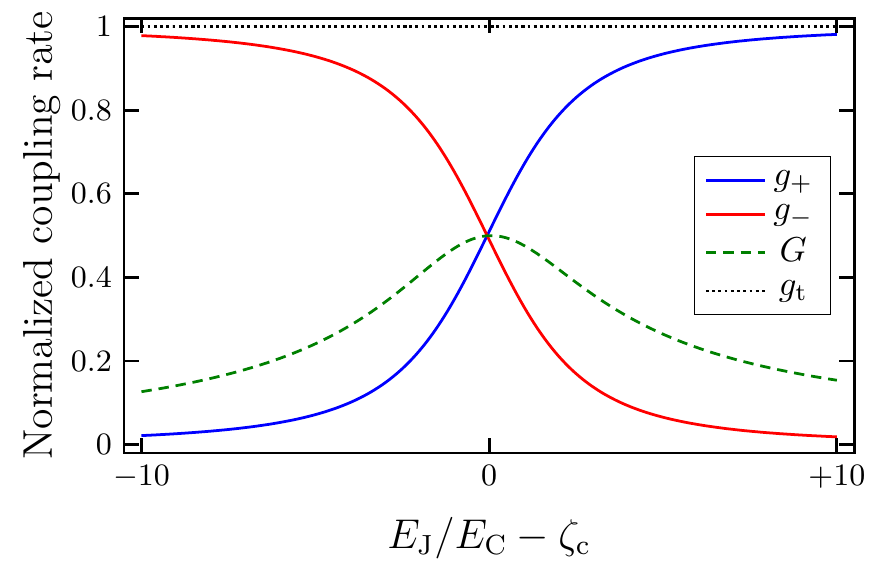}
\caption{Rates of the coupling between polaritons and the mechanical resonator normalized to the transmon--mechanical coupling rate $g_{\rm t}$.
}
\label{fig:Scoupling}
\end{figure}

To have an illustration of the coupling rates, we plot the polariton--mechanical coupling rates in Fig.~\ref{fig:Scoupling}.
The couplings $g_{\rm c}$ and $g_{\rm t\rm c}$ are not shown in the plot, because the corresponding lines almost coincide with the horizontal axis.
One notes that on-resonance all the three polariton--mechanical interactions are the same and equal to half a qubit--mechanics interaction.
We will use this property of the polaritonic interactions to prepare mechanical Fock states.
Also, for large cavity--transmon detunings the coupling rate of the three-mode interaction becomes negligible and can be disregarded in this regime.

%
%
%----------COOLING----------%
\section{Cooling the mechanical resonator}
\begin{figure}[b]
\includegraphics[width=0.6\linewidth]{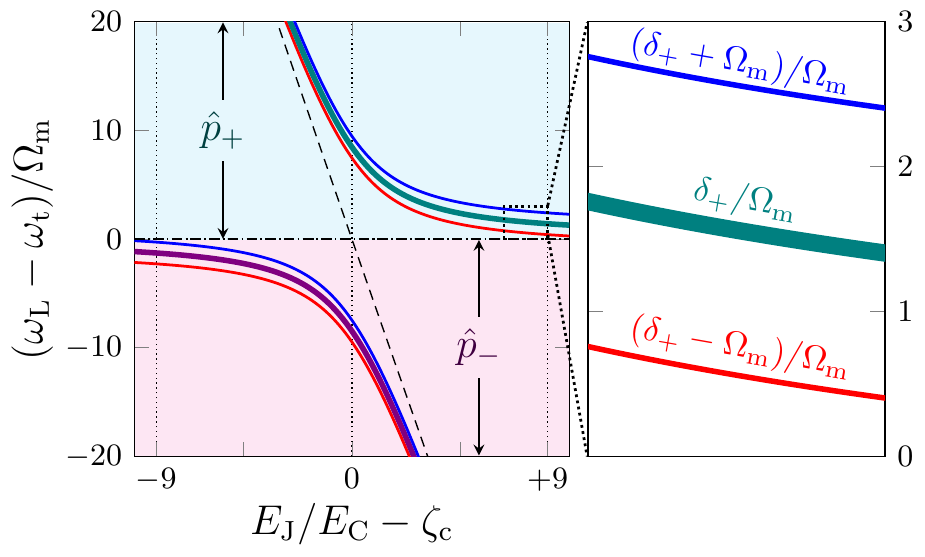}
\caption{(color online).
Avoided crossing and the blue and red sidebands of the polaritonic modes (green and violet bold lines) in the plane of $\Ejos/\Echr$ and the normalized detuning of the input drive with respect to the transmon frequency.
}
\label{fig:Ssideband}
\end{figure}

The interactions between the polaritons and the mechanical resonator make it possible to cool the mechanical resonator toward its ground state.
To this end we work in the regime where the inter-polariton interactions can be neglected via a rotating wave approximation.
This approximation will hold for $\omega_{+}-\omega_{-}=[\Delta^{2}+4\chi^{2}]^{\frac{1}{2}} \gg \Echr/2\hbar$.
For the parameters listed in Table~\ref{parameters} this approximation does hold only if $\Delta \gg \Echr/2\hbar$.
Working in the off-resonance regime means that only the qubit-like polariton is effectively interacting with the mechanical resonator.
This is because $g_{\rm c}$ and $g_{\rm t\rm c}$ are much smaller than $g_{\rm t}$, see Fig.~\ref{fig:Scoupling}.
In such situations, one only drives the \textit{interacting} polariton mode.
Therefore, when the cavity and the transmon are off resonance one arrives at the following Hamiltonian for the system in the frame rotating at the frequency of the input drive $\omega_{\rm L}$:
\begin{eqnarray}
\frac{\Hh'}{\hbar} &=&\wm\bhd\bh
-\delta_{+}\ph_{+}^{\dagger}\ph_{+} -\lambda_{+}\ph_{+}^{\dagger}\ph_{+}^{\dagger}\ph_{+}\ph_{+} 
-\delta_{-}\ph_{-}^{\dagger}\ph_{-} -\lambda_{-}\ph_{-}^{\dagger}\ph_{-}^{\dagger}\ph_{-}\ph_{-} \nonumber\\
&&+(g_{+}\ph_{+}^{\dagger}\ph_{+} +g_{-}\ph_{-}^{\dagger}\ph_{-})(\bh+\bhd)
+\Lambda_{0} ~\ph_{+}^{\dagger}\ph_{+}\ph_{-}^{\dagger}\ph_{-} \nonumber\\
&&-\alpha_{-}\Dlasr(\ph_{+}+\ph_{+}^{\dagger})
+\alpha_{+}\Dlasr(\ph_{-}+\ph_{-}^{\dagger}).
\end{eqnarray}
Here, $\hbar\Lambda_{0}=2\Echr\alpha_{+}^{2}\alpha_{-}^{2}$ is the polaritonic intensity--intensity interaction and $\delta_{\pm}=\omega_{\rm L}-\omega_{\pm}$ is the detuning of the polariton.
Out of resonance, only one of the polaritons is `truly' fed by the input drive, because optimal sideband cooling happens for $\delta_{\pm}\approx -\wm$ and this leads to $\delta_{\pm} \gg \delta_{\mp}$ making one of the polaritons an idler, see Fig.~\ref{fig:Ssideband}.

\begin{table}[t]
\caption{\label{parameters}Parameters of the system.}
\begin{ruledtabular}
\begin{tabular}{lll}
Quantity & set\#1 & set\#2 \\
\colrule
$\zeta$ & 150\footnote{resonance value, i.e., $\wt(\zeta)=\wc$.}$\pm$50 & 142$^{\rm a}$$\pm$60 \\
$m$ & 1 pg & 3 pg \\
$g_{0}$ & 18.2 kHz & 20.6 kHz \\
$g_{\rm t}(\zeta_{\rm c})$ & 315 kHz & 350 kHz \\
$\chi(\zeta_{\rm c})$ & 31.4 MHz & 510 MHz \\
$\wm/2\pi$ & 10 MHz & 1 MHz \\
$\kappa_{\rm c}/2\pi$ & 10 kHz & 50 kHz \\
$\gamma_{\rm t}/2\pi$ & 3 kHz & 5 kHz \\
$8\Echr n_{\rm{ac}}/\hbar\wc$\footnote{prefactor of the qubit--cavity coupling rate whose fundamental \textit{upper} bound is $\sim 0.1$.} & $2.0\times 10^{-4}$ & $3.4\times 10^{-3}$
\end{tabular}
\end{ruledtabular}
\end{table}

%
%
%----------PROTOCOL----------%
\section{State preparation protocols}

%----------FOCK----------%
\subsection{Mechanical Fock states}
Here we employ the three-mode interaction to convey the excitations to the mechanical resonator.
According to Fig.~\ref{fig:Scoupling} the regime in which such interactions are significant is when the cavity and the transmon are in resonance ($\wt \approx \wc$).
Moreover, one needs to have $|\omega_{+}-\omega_{-}|\approx\wm$ to make the state transfer.
Specifically, we consider the case of $\omega_{+}-\omega_{-}=\wm$ which for around resonance means that $\chi \approx \wm/2$ must hold, see Fig.~\ref{fig:Sprotocol}(a).

The following recipe can be used to prepare Fock states in the mechanical resonator:
(i) Prepare the system in its ground state: $\ket{0_{+},0_{-},0_{\rm m}}$.
(ii) Flip up the `+' polariton by applying a $[\pi]_{+}$ pulse to get $\ket{1_{+},0_{-},0_{\rm m}}$.
(iii) Let the system evolve for a duration of $\tau_{1}=\pi/2G $.
Assuming fast enough single qubit gate operations this leads to the state $\ket{0_{+},1_{-},1_{\rm m}}$.
Note that basically the inter-polariton interactions cannot affect this process because from Eq.~(\ref{Spp}) it can be easily proven that the sates $\ket{0_{+},1_{-}}$ and $\ket{1_{+},0_{-}}$ do not change under such interactions.
However, to produce higher Fock states one needs to get rid of the excitation in the `$-$' polariton mode.
Hence, the fourth step of the protocol must be:
(iv) By applying a $[\pi]_{-}$ pulse the `$-$' polariton is flipped down $\ket{0_{+},0_{-},1_{\rm m}}$.
To reach the target state:
(v) Repeat steps (ii) to (iv) by replacing $\tau_{1}$ with $\tau_{n}=\pi/2\sqrt{n}G$, where $n$ is the number of the rounds the protocol has been repeated (or equivalently, the number of phonons in the next Fock state).
To detect or verify this state, one could reverse the steps and measure state of the qubit at the end.

%----------ENTANGLED----------%
\subsection{Tripartite hybrid entanglement}
\begin{figure}
\includegraphics[width=0.7\linewidth]{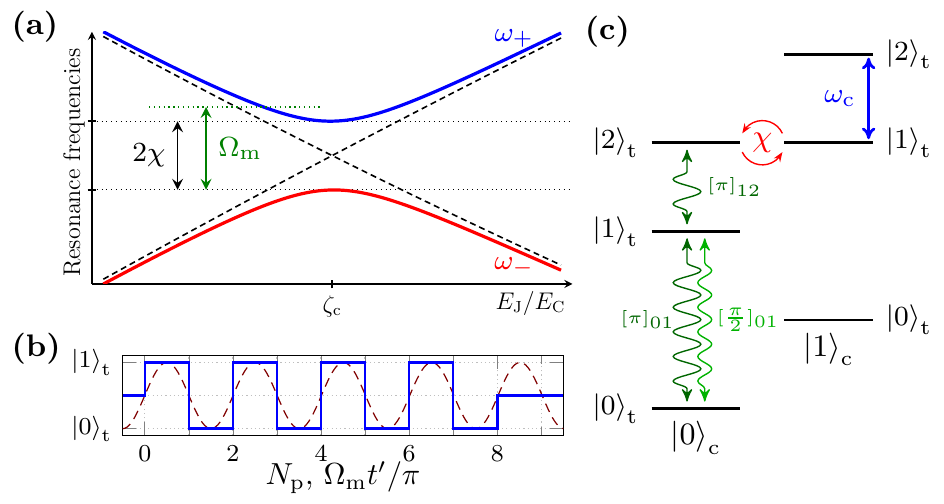}
\caption{
(a) Resonance frequencies of the system with the dressed resonance frequencies appropriate for the preparation of mechanical Fock states.
(b) Pulse sequence applied to the qubit for maximally displacing the mechanical resonator with 7 flips (solid line). The oscillation of the mechanical resonator is displayed for comparison (dashed line).
(c) Energy level diagram of the transmon when on-resonance with the cavity, suited for generating the tripartite entangled state.
}
\label{fig:Sprotocol}
\end{figure}

A tripartite entanglement between components can be prepared in the considered system.
We work with the original (not the dressed) picture of the system to describe the steps for creating such states.
We also consider an effective three-level model for the qubit with $\ket{0}_{\rm t}$ its ground state, $\ket{1}_{\rm t}$ and $\ket{2}_{\rm t}$ its first and second excited states.
Furthermore, because the protocol is mostly executed in the off-resonance regime, the qubit-mechanical interaction is the only important interaction of the mechanical resonator while the optomechanical and three-mode interactions are negligible.
We assume that the cavity is fabricated such that the frequency of its fundamental mode matches $\omega_{12}$ of the transmon, that is, when on resonance the cavity shares excitations with the first and second excited states of the qubit, see Fig.~\ref{fig:Sprotocol}(c).

The steps for preparing the state are the following:
(i) When the qubit and the cavity are off resonance, the system is prepared in its ground state $\ket{0_{\rm t},0_{\rm c},0_{\rm m}}$.
(ii) A $[\pi/2]$ pulse is applied to the qubit to make the superposition state $\frac{1}{\sqrt{2}}\big(\ket{0_{\rm t},0_{\rm c},0_{\rm m}}+\ket{1_{\rm t},0_{\rm c},0_{\rm m}}\big)$.
This superposition state in the qubit results in a state dependent force on the mechanical resonator such that when the qubit is in its ground state the mechanical resonator will experience no force, and therefore no displacement and relevant phase shift, while when the qubit is excited a force will be applied to the mechanics.
(iii) Let the mechanics and the transmon interact for a time interval $t^{\prime}$.
Omitting a global irrelevant phase factor one gets $\frac{1}{\sqrt{2}}\big[\ket{0_{\rm t},0_{\rm c},0_{\rm m}}+e^{i\phi(t^{\prime})}\ket{1_{\rm t},0_{\rm c},\alpha(t^{\prime})}\big]$, where the mechanical part is in a coherent state with $\alpha(t^{\prime})=\frac{g_{\rm t}}{\wm}(e^{-i\wm t^{\prime}}-1)$ and the phase difference is $\phi(t^{\prime})=\frac{g_{\rm t}^{2}}{\wm^{2}}(\wm t^{\prime}-\sin\wm t^{\prime})$.
The relation for $\alpha(t^{\prime})$ already shows that since $g_{\rm t} \ll \wm$ it is impossible to have a superposition of macroscopically distinguishable mechanical states after this single preparation step.
However, it is possible to increase the `distance' between the two mechanical states by applying a sequence of $[\pi]$ pulses at time intervals equal to one half mechanical period ($t^{\prime}=\pi/\wm$) [Fig.~\ref{fig:Sprotocol}(b)].
In this way, by applying $N_{\rm p}$ of such pulses we arrive at $\frac{1}{\sqrt{2}}\big[\ket{0_{\rm t},0_{\rm c},-\beta_{0}}+e^{-i\theta}\ket{1_{\rm t},0_{\rm c},\beta_{1}}\big]$ where
\begin{equation}
\beta_{0}-\beta_{1}=2(N_{\rm p}+1)\frac{g_{\rm t}}{\wm}
\end{equation}
is the distance between the coherent states and $\theta$ is the relative phase, which is zero for odd number of pulses and equals to $\theta=\pi(\wt -\frac{g_{\rm t}^{2}}{\wm})/\wm$ for even $N_{\rm p}$.
In order to have a symmetric displacement around origin of the phase space, it is reasonable to choose an odd number of such pulses $N_{\rm p}$ which results in $\frac{1}{\sqrt{2}}\big[\ket{0_{\rm t},0_{\rm c},\beta}+\ket{1_{\rm t},0_{\rm c},-\beta}\big]$ with $\beta\equiv (N_{\rm p}+1)g_{\rm t}/\wm$.
(iv) By applying another $[\pi/2]$ pulse to the transmon, we push the state into a superposition of even and odd cat mechanical states $\frac{1}{2}\big(\ket{0_{\rm t},0_{\rm c},\psi_{+}}+\ket{1_{\rm t},0_{\rm c},\psi_{-}}\big)$, where $\mathcal{N}_{\pm}\psi_{\pm}=\mathcal{N}_{\pm}\big(\ket{\beta}\pm \ket{-\beta}\big)$  is an even or odd cat state with the normalisation factor $\mathcal{N}_{\pm}=\big[2\pm 2e^{-2|\beta|^{2}}\big]^{-\frac{1}{2}}$.
(v) A $[\pi]_{1\leftrightarrow 2}$ pulse flips the qubit $\ket{1}_{\rm t} \rightarrow \ket{2}_{\rm t}$.
(vi) A flux pulse of duration $\pi/2\sqrt{2}\chi$ brings the relevant transition of the transmon into resonance with the cavity ($\omega_{12}=\wc$) and transfers the qubit \textit{second} excitations to the cavity. Finally we thus arrive at the following state:
\begin{equation}
\frac{1}{2}\big(\ket{0_{\rm t},0_{\rm c},\psi_{+}}+\ket{1_{\rm t},1_{\rm c},\psi_{-}}\big).
\end{equation}
This is a hybrid Greenberger--Horne--Zeilinger state.

%\end{widetext}
%\bibliography{transmon}

\end{document}